\documentclass[prd,twocolumn,superscriptaddress,floatfix,nopacs,preprintnumbers,nofootinbib,10pt,aps]{revtex4-2}
\usepackage[utf8]{inputenc}

\usepackage[normalem]{ulem}
\usepackage{xcolor}
\usepackage{mathtools}
\usepackage[breaklinks,colorlinks,citecolor=citcolor,urlcolor=blue,linkcolor=lcolor]{hyperref}

\usepackage{footnote}
\usepackage{mathrsfs}
\usepackage{physics}
\usepackage{amssymb,bm}
\usepackage{amsmath}
\usepackage{slashed}
\usepackage{graphics}
\usepackage{graphicx}
\usepackage{dsfont}

\newcommand{\Nc}{N_\mathrm{c}}

\newcommand{\bt}{\boldsymbol{b}_\perp}
\newcommand{\rt}{\boldsymbol{r}_\perp}
\newcommand{\qt}{\mathbf{q_\perp}}

\newcommand{\pt}{\boldsymbol{p}_{\perp}}
\newcommand{\xt}{\boldsymbol{x}_\perp}
\newcommand{\Bt}{\boldsymbol{B}_\perp}

\newcommand{\yt}{\boldsymbol{y}_\perp}

\newcommand{\as}{\alpha_\mathrm{s}}
\newcommand{\der}{\mathrm{d}}
\newcommand{\xq}{\boldsymbol{x}_{0\perp}}
\newcommand{\xaq}{\boldsymbol{x}_{1\perp}}

\newcommand{\dm}{\mathrm{D}^0}
\newcommand{\dme}{\mathrm{D}}

\newcommand{\kq}{\boldsymbol{k}_{0\perp}}
\newcommand{\kqm}{k_{0\perp}}
\newcommand{\kd}{\boldsymbol{k}_{\mathrm{D}\perp}}
\newcommand{\kdm}{k_{\mathrm{D}\perp}}
\newcommand{\kaq}{\boldsymbol{k}_{1\perp}}
\newcommand{\lt}{\boldsymbol{\ell}_\perp}
\newcommand{\ltm}{\ell_\perp}

\newcommand{\aem}{\alpha_{\mathrm{e.m.}}}
\newcommand{\Qs}{Q_\mathrm{s}}

\definecolor{lcolor}{rgb}{0.5,0,0}
\definecolor{citcolor}{rgb}{0,0.3,0.0}
\usepackage{multirow}
\usepackage{booktabs}  
\usepackage{siunitx}

\begin{document}

\author{Patricia Gimeno-Estivill}
\email{patricia.p.gimenoestivill@jyu.fi}
\author{Tuomas Lappi}
\email{tuomas.v.v.lappi@jyu.fi}
\author{Heikki Mäntysaari}
\email{heikki.mantysaari@jyu.fi}
\affiliation{
Department of Physics, University of Jyväskylä,  P.O. Box 35, 40014 University of Jyväskylä, Finland
}
\affiliation{
Helsinki Institute of Physics, P.O. Box 64, 00014 University of Helsinki, Finland
}

\title{Inclusive \texorpdfstring{$\mathrm{D}^0$}{D0} Photoproduction in Ultraperipheral Collisions }

\begin{abstract}
We compute the differential cross section for inclusive $\dm$ production in ultraperipheral collisions within the Color Glass Condensate framework. Our predictions are found to be in relatively good agreement with the CMS data at small transverse momentum, which is the region of validity of our approach. Furthermore, we quantify saturation effects by a nuclear modification ratio $R_{pA}$ for $\dm$ photoproduction and examine both analytically and numerically the collinear factorization limit of the $\dm$ differential photoproduction cross section.
\end{abstract}

\maketitle 

\section{Introduction}
In ultraperipheral heavy-ion  collisions (UPCs), the gluonic structure of the nucleus is probed by a quasireal photon~\cite{Bertulani:2005ru,Klein:2019qfb}. Since hadronic interactions are suppressed, UPCs are clean processes to detect experimentally.
In recent years there have been many measurements of exclusive vector meson photoproduction in ultraperipheral proton-lead and lead-lead  collisions at RHIC~\cite{STAR:2023nos,STAR:2023vvb} and at the LHC  \cite{ALICE:2023jgu, ALICE:2023mfc, CMS:2023snh,ALICE:2020ugp,ALICE:2023kgv,LHCb:2022ahs,LHCb:2018rcm}. Lately, the UPC program has broadened to inclusive measurements, starting from dijet production~\cite{ATLAS:2024mvt}. Very recently inclusive $\dm$ data has been collected~\cite{CMS}  for the first time. The purpose of this paper is to address this measurement from a small-$x$ and saturation point of view.

In this work, we calculate inclusive $\dm$ photoproduction cross section in lead-lead UPCs in the CGC picture. Concretely, we focus on the ``Xn0n'' event selection, where the projectile nucleus remains intact while the target nucleus undergoes nuclear breakup, emitting ``X'' neutrons after the interaction. These neutrons are experimentally detected using zero-degree calorimeters (ZDCs). Within this setup, we compare our theoretical predictions to preliminary CMS data on inclusive $\dm$ photoproduction \cite{CMS}. 
Our result for massive quark-pair production is in agreement with Ref.~\cite{Dominguez:2011wm}, and our expression for inclusive charm production is consistent with Refs.~\cite{Goncalves:2017zdx, Marquet:2009ca}. 

At the high energies accessible at the LHC, photon-nucleus scattering allows for the study of gluon saturation in nuclei at Bjorken-$x$ values down to $10^{-6}$ \cite{Khatun:2024vgn}. To quantify nuclear effects, we introduce a nuclear modification ratio $R_{pA}$ for $\dm$ photoproduction and also investigate the collinear factorization limit of the $\dm$ differential cross section. Although the small-$x$ values probed at the LHC are beyond the reach of the future Electron-Ion Collider (EIC), our study can lay the base for future  inclusive measurements in electron-nucleus scattering at the EIC.  

The paper is organized as follows. In Sec.~\ref{Sec: Theoretical framework} we derive the inclusive $\dm$ photoproduction cross section in the CGC and compute its collinear factorization limit.  
The numerical study is presented in Sec.~\ref{sec:results}, where we compare our predictions with preliminary CMS data and illustrate the collinear factorization expression as well as the nuclear modification factor $R_{pA}$. We conclude with a summary in Sec.~\ref{sec: summary}.

\begin{figure}[tb]
    \centering
    \includegraphics[width=0.9\linewidth]{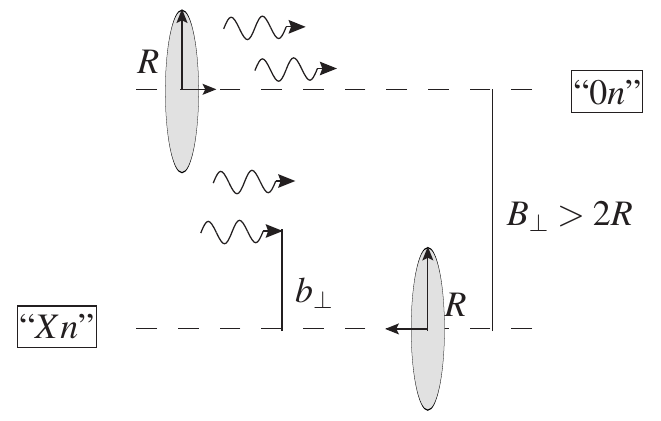}
    \caption{Ultraperipheral nucleus-nucleus collision where a flux of photons is coherently emitted by the Lorentz-contracted right-moving nucleus. Here $B_\perp$ is the impact parameter and $b_\perp$ represents the transverse distance between the quasi-real photon probe and the center of the target nucleus, which moves to the left. After the interaction, the emitted ``X'' neutrons (at least one) from the target in its forward direction ($y<0$) are  detected in zero-degree calorimeters. }
    \label{fig: UPC}
\end{figure}

\section{Theoretical framework}
\label{Sec: Theoretical framework}
In this section we will derive the differential cross section for inclusive hadron photoproduction in UPCs using the CGC formalism and including quark masses. 
As illustrated in Fig.~\ref{fig: UPC}, UPCs occur when the distance $B_\perp$ between the centers of the two nuclei is greater than twice their radius  $R=(1.2A^{1/3}-0.86A^{-1/3} )\; \mathrm{fm}^2$, where $A$ is the atomic number.  At this separation, hadronic interactions are suppressed and the interaction is mediated by a photon.
We will start by writing down the expression of the photon flux that mediates the interaction. Then, using Light-Cone Perturbation Theory and the Color Glass Condensate~\cite{Iancu:2003xm}, we will calculate the partonic cross section representing the photon splitting into a quark pair and the subsequent interaction with the target nucleus. Finally, we will convolute our expression with a fragmentation function to obtain the $\dm$ differential cross section. The final result is written in Eq.~(\ref{eq: result cross section}). 
\subsection{Photon flux}
\label{subsec:UPC}
At high energies, the electromagnetic field emitted by a point-like particle is Lorentz contracted in the direction parallel to its motion. As a result, the interaction time between the transversely boosted electromagnetic field and a target becomes very short: $\Delta t=B_\perp/\gamma$, where $B_\perp$ is the impact parameter and $\gamma$ the Lorentz factor. This interaction can be approximated as a real pulse of light incident on the target in the Weizsäcker-Williams method (one-photon exchange) \cite{Williams:1934ad,vonWeizsacker:1934nji}. In this approximation,  the Fourier transform of the Lorentz-contracted electromagnetic fields emitted by a high-energy charged particle is equivalent to a flux of photons $N(\omega)$ with energy $\omega$ \cite{Bertulani:2005ru,Baur:1998ay}:
\begin{equation}
\begin{aligned}
\label{eq:photon_flux}
    N(\omega)&=\int_{B_\perp>2R}\mathrm{d}^2\Bt\frac{Z^2\aem\omega^2}{\pi^2\gamma^2}K_1^2\bigg(\frac{\omega B_\perp}{\gamma}\bigg) \;.
\end{aligned}
\end{equation}
Here $Z$ is the ion charge, $\aem$=1/137 the fine structure constant and $\gamma=A\sqrt{s}/(2M_A)$ is the boost factor of the nucleus, with $\sqrt{s}$ the nucleon-nucleon center-of-mass collision energy. We consider lead nuclei with charge $Z=82$, atomic number $A=208$ and mass $M_A=193.64$ GeV. The modified Bessel function $K_1$ gives the flux for transversely polarized photons.
The photon flux is large in the regime when they are coherently emitted from the whole nucleus, which leads to the coherence condition~\cite{Baur:2001jj}:
\begin{equation}
    Q^2\lesssim\frac{1}{R^2} \;.
\end{equation}
Because the nuclear radius is large,
the photons can be considered as quasi-real, with a maximum value of $Q^2_{\mathrm{max}}\approx 0.001 \; \mathrm{GeV}^2$ for a lead nucleus.

Thus, the UPC is mediated by a quasi-real photon from one of the swarms of equivalent photons  that ``interacts directly''\footnote{The photon is colorless, it splits into a quark pair that then interacts with the gluon field in the target nucleus.} with the target nucleus. As it is depicted in Fig.~\ref{fig: UPC}, we concretely choose our coordinate system such that the right-moving nucleus is the photon source and the left-moving nucleus the target. 
We will want to contrast our calculation with CMS data taken with an ``Xn0n'' event selection, referring to at least one neutron (``Xn'') being detected on one side and no neutrons (``0n'') on the other side by zero-degree calorimeters  \cite{Krauss:1997vr}. Since we are looking at inclusive photon-target scattering with no rapidity gap between the produced system and the target, the target nucleus will practically always break up and emit neutrons, and we do not need to specifically include the ``Xn'' condition in our calculation. The photon-emitting nucleus, on the other hand, can break up due to Coulomb dissociation, emit neutrons and fall outside of the ``0n'' category.
To incorporate this condition in our  predictions, we include in the cross section  a factor $P(B_\perp)$, which represents the probability that the projectile nucleus does not emit any neutrons (0n) using the parametrization introduced in Ref.~\cite{Baur:1998ay}:
\begin{equation}
    P(B_\perp)=\mathrm{exp}(-S/B_\perp^2) \;,
    \label{eq:0n_emission}
\end{equation}
where
\begin{equation}
    S=5.45 \cdot 10^{-5}\frac{Z^3(A-Z)}{A^{2/3}} \;.
\end{equation}
For lead nucleus $S=(10.4\;\mathrm{fm})^2$.  
In our numerical analysis, we also use the parameter $S=(17.4\;\mathrm{fm})^2$ obtained in Refs. \cite{Baltz:1997di,Baltz:1996as}. The two different values applied here enable us to estimate the uncertainty related to the description of the electromagnetic dissociation process.

Integrating the photon flux in Eq.~\eqref{eq:photon_flux} over the photon energies $\omega$, convoluting with the parton level cross section and including the ``0n'' emission probability Eq.~\eqref{eq:0n_emission}, the cross section for inclusive $\dm$ meson photoproduction with transverse momentum $\kd$ and rapidity $y_{\mathrm{D}}$ in UPC can be expressed as
\begin{widetext}
\begin{equation}
\label{eq: convolution}
\frac{\mathrm{d}\sigma^{A+A \rightarrow \dm +X}}{\der^2\kd \der y_{\mathrm{D}}} = {\int\der \omega\int_{B_\perp > 2R}  \der^2\Bt P(B_\perp)F(\omega, B_\perp)}  
     \int\der z_f{ \frac{\mathrm{d}\sigma^{\gamma+A \rightarrow c+X}}{\der^2\kq \der y_0}}
      M (z_f, m) \;.
\end{equation}
\end{widetext}
Here, the $\sigma^{\gamma+A \rightarrow c+X}$ is the partonic cross section to produce a charm quark with transverse momentum $\kq=\kd/z_f$ and rapidity $y_0$ from a photon-nucleus scattering.
We derive its explicit expression using Light-Front Pertubration Theory in the next subsection~\ref{subsec: Light-Front quantization}. The fragmentation function $M(z_f,m)$ in Eq.~\eqref{eq: convolution} represents the hadronization of the charm quark with mass $m\sim m_\mathrm{D}$ into the $\dm$ meson. 
As a non-perturbative function, we use the parametrization obtained in Ref.~\cite{Kniehl:2006mw}, which we write in Eq.~\eqref{eq: FF} in subsection~\ref{sub: final result}.

\subsection{Partonic cross section in Light-Cone Perturbation Theory}
\label{subsec: Light-Front quantization}
In this subsection, we first derive the partonic cross section in LCPT and CGC formalism for the leading order quark-pair production process: 
\begin{equation}
\label{eq: process}
    \gamma_\lambda  + A \rightarrow q\Bar{q} + X \;.
\end{equation} 

Subsequently, we will integrate over the momentum of the antiquark to obtain the inclusive charm production cross section in the CGC formalism. Our expressions are consistent with the results for massless and also heavy quarks obtained in Refs.~\cite{Dominguez:2011wm, Marquet:2009ca, Goncalves:2017zdx}. The mathematical conventions used for the normalization of the states and the Fourier transforms are given in Appendix~\ref{sec: Appendix}. 

\begin{figure}[tb]
    \centering
    \includegraphics[width=0.85\linewidth]{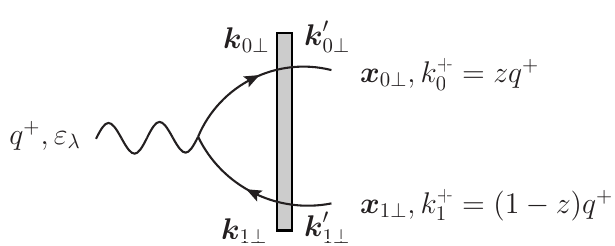}
    \caption{The quasi-real photon with $\boldsymbol{q}_\perp=0$ splits into a quark pair that interacts with the gluon shockwave in the target nucleus, represented by a gray band. The transverse momentum of the quark(antiquark) is $\kq(\kaq)$ before the interaction and $\kq'(\kaq')$ after it. The quark transverse positions remains fixed at $\xq (\xaq)$ throughout the interaction (\textit{eikonal} scattering). The quark carries $k_0^+=zq^+$ of the longitudinal momentum of the photon, while the antiquark takes $k_1^+=(1-z)q^+$. }
    \label{fig: LO splitting}
\end{figure}

We work in light-cone coordinates where $v^\pm=(v^0 \pm v^3)/\sqrt{2}$ and $\boldsymbol{v}_\perp=(v^1,v^2)$ for any 4-vector $v^\mu$. In this notation, the momentum of the incoming photon is expressed as $q^\mu=(q^+, 0^-,\boldsymbol{0}_\perp)$ and its transverse polarization by $\epsilon^\mu_\lambda= (0^+,0^-,\boldsymbol{\varepsilon}_\lambda)$. Here $\boldsymbol{\varepsilon}_{\pm} $ are circular polarization vectors, written in Eq.~\eqref{eq: Circular polarization operators}. Before the interaction, the photon splits into a quark and antiquark with respectively momenta $k^\mu_0=(zq^+,k_2^-,\kq)$ and  $k^\mu_1=((1-z)q^+,k_1^-,\kaq)$, where $\kq=-\kaq$. The $z=k_0^+/q^+$ is the fraction of longitudinal momentum transferred from the photon to the quark. The colored partons then interact with the target nucleus which is described as a classical gluon field $A^-(x)$ in the Color Glass Condensate. As illustrated in Fig.~\ref{fig: LO splitting} and discussed later in this subsection, the target does not contribute any $k^+$ momentum in the interaction. 

To obtain the invariant scattering amplitude $\mathcal{M}^{\gamma\rightarrow q\Bar{q}}$ for the process in Eq.~\eqref{eq: process}, we begin by expanding the physical state of the quasi-real incoming photon over a basis of Fock states in \textit{momentum} space \cite{Beuf:2022kyp}: 
\begin{equation}
\label{eq: fock state expansion}
\begin{aligned}
    \big|\gamma_\lambda(q^+,0^-,\boldsymbol{0}_\perp)\big\rangle_{\mathrm{D}} &= \sqrt{Z_{\gamma_\lambda}}\Bigg\{ \sum_{q_0\Bar{q}_1}\Psi ^{\gamma_\lambda \rightarrow q_0\Bar{q}_1}|q_0\Bar{q}_1\rangle_0 \\ &+ \sum_{q_0\Bar{q}_1g_2}\Psi ^{\gamma \rightarrow q_0\Bar{q}_1g_2}|q_0\Bar{q}_1g_2\rangle_0 + \dots \Bigg\} \; \\
    &=  \sum_{q_0\Bar{q}_1}\Psi ^{\gamma_\lambda \rightarrow q_0\Bar{q}_1}b^\dagger_0d^\dagger_1 |0\rangle  + \dots\;.
\end{aligned}
\end{equation}
Here $|\gamma_\lambda(q^+,0^-,\boldsymbol{0}_\perp)\rangle_{\mathrm{D}}$ is the physical (dressed) state in the interaction picture, while the free (bare) states are denoted by the subscript $0$. We work here at leading order in both the strong and electromagnetic couplings. This implies that we use a  one-photon exchange approximation and neglect higher order contributions in  the renormalization factor $Z_{\gamma_\lambda^*}=1+\mathcal{O}(e^2)$. Additionally, at leading order in the strong coupling constant $\alpha_s$, we neglect higher order bare states that involve additional gluons. The notation $\sum_{q_0\Bar{q}_1}$ represents a sum over the quantum numbers of each parton in the Fock state and it includes a phase-space integration \cite{Kovchegov:2012mbw}:
\begin{equation}
\label{eq: sum}
\sum_{q_0\Bar{q}_1}:=\sum_{h_0,\alpha_0,f_0}\sum_{h_1,\alpha_1,f_1}\prod_{i=0}^1\bigg[ \int \frac{\der k_i^+}{(2\pi)}\frac{\theta(k_i^+)}{2k_i^+}\frac{\der^2\boldsymbol{k}_{i \perp}}{(2\pi)^2}\bigg] \;.
\end{equation}
Here $\alpha_{0}(\alpha_1)$ represents the color of the quark(antiquark) and $h_{0}(h_1)$ their helicity ($h_i=\pm \frac{1}{2}$). For the quark and antiquark creation operators: $b_0^\dagger$ and $d_1^\dagger$, we use the shorthand notation $b^\dagger_0 \coloneqq b(k_0^+,\kq,h_0,\alpha_0)$ and $d^\dagger_1\coloneqq d(k_1^+,\kaq,h_1,\alpha_1)$. The light-cone wave functions (LCWFs) $\Psi$ in Eq.~\eqref{eq: fock state expansion} can be computed perturbatively from interaction vertices. These conserve the spatial momentum $\underline{p}\equiv(p^+,\boldsymbol{p}_\perp)$ but not the light-cone energy $p^-=(m^2+\pt^2)/2p^+$. This is because we are working in ``old-fashioned'' perturbation theory, where the perturbative expression is expressed in terms of on-shell energies of intermediate particles, but interactions cause transitions between these different energy bare states. 

At the lowest order in $\as$, the LCWF in Eq.~\eqref{eq: fock state expansion} is 
\begin{equation}
\label{eq: LCWF}
\begin{aligned}
    \Psi^{\gamma_\lambda \rightarrow q_0\Bar{q}_1}(\kq,\kaq,z)&=\frac{\langle 0|d_1b_0\hat{V}_I(0)a_\gamma^\dagger|0\rangle}{\mathrm{ED}} \;,
\end{aligned}
\end{equation}
where $\hat{V}_{I}(0)$ is the interaction operator and ED is the energy denominator that contain the light-front energies of the incoming and outgoing states:
\begin{equation}
\label{eq: sol ED}
\begin{aligned}
    \mathrm{ED}& = q_{\gamma}^- - k_{q_0\Bar{q}_1}^- \\
    &= \frac{\boldsymbol{q}_\perp^2 - Q^2}{2q^+} - \frac{\kq^2 + m^2}{2k_0^+} - \frac{\kaq^2 + m^2}{2k_1^+} \\
    &= -\frac{(\kq^2 + m^2)}{2q^+ z (1 - z)} \;.
\end{aligned}
\end{equation}
Here $Q^2=0$ and $\boldsymbol{q}_\perp=0$ for the quasi-real photon. We have also used the momentum conservation $q^+=k_0^++k_1^+$ and expressed longitudinal momenta in terms of the momentum fraction $z=k_0^+/q^+$.
These light-cone energy denominators, calculated with on-shell single particle energies, represent the off-shellness of the whole multiparticle intermediate state. They correspond to propagators, which in covariant theory  are forced to be off-shell by energy  conservation at the vertices.

The expression for the interaction operator $\hat{V}_{I}$ and the calculation of the numerator in Eq.~\eqref{eq: LCWF} can be found, e.g., in Refs.~\cite{Brodsky:1997de,Beuf:2016wdz,Kovchegov:2012mbw}. The LCWF in Eq.~\eqref{eq: LCWF} can then be written as an interaction vertex with a momentum conservation delta function: 
\begin{align}
 \label{eq: LCWF 2}
&    \Psi(\kq,\kaq,z) \nonumber \\ 
&   \quad =(2\pi)^3\delta^3(\underline{q}-\underline{k_1}-\underline{k_0})\delta_{\alpha_0\alpha_1}\frac{ee_f\Bar{u}_0\slashed{\epsilon}_\lambda(q^\mu)v_1}{\mathrm{ED}}\\ \nonumber
 &  \quad  =-2q^+(2\pi)\delta(q^+-k_1^+ -k_0^+)\psi(\kq,\kaq,z) \:.
\end{align}
In this expression $e$ is the electromagnetic constant and $e_f$ the fractional charge of the quark: $e_c=2/3$ for charm and $e_b=-1/3$ for bottom quarks. The longitudinal momentum $2q^+$ in the second line arises from the energy denominator in Eq.~\eqref{eq: sol ED}. By factorizing out the longitudinal momentum conservation, we introduced the \textit{reduced wave function} $\psi(\kq,\kaq,z)$.

The matrix elements between the spinors in Eq.~\eqref{eq: LCWF 2} are tabulated, see e.g. Refs.~\cite{Hanninen:2017ddy,Kovchegov:2012mbw}. For transversally polarized photons they take the form:
\begin{widetext}
\label{eq: vertex spinors}
\begin{equation}
    \Bar{u}_{0}\slashed{\epsilon}_\lambda(q^\mu)v_1=\bigg[\delta_{h_0,-h_1}\frac{2}{\sqrt{z(1-z)}}\Big(z\delta_{\lambda,2h_0}-(1-z)\delta_{\lambda,-2h_0}\Big)\varepsilon_\lambda\cdot\kq+ \delta_{h_0,h_1}\delta_{\lambda,2h_1}\frac{\sqrt{2}m}{\sqrt{z(1-z)}}\bigg] \;.
\end{equation}
\end{widetext}
This expression separates the helicity-conserving vertex, which is independent of the mass, from the helicity-flip matrix element, which is proportional to the mass.

With the explicit form of the physical incoming state in Eq.~\eqref{eq: fock state expansion}, we can now calculate the invariant scattering amplitude $\mathcal{M}^{\gamma\rightarrow q\Bar{q}}$ from the matrix elements of the incoming and outgoing states \cite{Bjorken:1970ah,Beuf:2022kyp,Hanninen:2021byo}:
\begin{multline}
\label{eq: relation}
    2q^+(2\pi)\delta(q^+-{k'_0}^+-{k'_1}^+)i\mathcal{M}(\qt=0\rightarrow {\kq}',{\kaq}') \\ 
    \equiv \big\langle q({\kq}',{k'_0}^{+})\Bar{q}({\kaq}',{k'_1}^{+})\big|(\hat{S}-\mathds{1})\big|
    \gamma(q^+,0^-,\boldsymbol{0}_\perp)\big\rangle \;.
\end{multline}
Here we subtract the identity case where there is no scattering. The operator $\hat{S}$ represents the \textit{eikonal} scattering of the incoming quark-pair with the gluon field in the target nucleus. In this eikonal interaction, the partons undergo a color rotation and transverse momentum is transferred from the target, but their helicities and transverse positions are conserved. The scattering operator $\hat{S}$ operates on the bare states in \textit{mixed} space as: 
\begin{equation}
\label{eq: color rotation}
    \hat{S}\big|q(q^+,\boldsymbol{x}_\perp,i)\big\rangle = \sum_j [U_F(\boldsymbol{x}_\perp)]_{ji} \big|(q^+,\boldsymbol{x}_\perp,j)\big\rangle \:,
\end{equation}
where $i,j$ denote the color indices of the quark and $\hat{S}|0\rangle = |0\rangle$.
The Wilson line operator $U_F(\boldsymbol{x}_\perp)$ in the fundamental representation represents the quark propagator in the gluon color field $A_a^-(x)$ of the target nucleus. It is expressed as \cite{Kovchegov:2012mbw} 
\begin{equation}
\label{eq: Wilson Line}
    U_F(\boldsymbol{x}_\perp)=\mathcal{P}\;\mathrm{exp}\Bigg(-ig\int \der x^+ A^-_a(x^+,\boldsymbol{x}_\perp) t^a \Bigg) \;,                
\end{equation}
where $\mathcal{P}$ is a path-ordering operator and $t^a$ fundamental color matrices. 

Since the scattering operator $\hat{S}$ acts in coordinate space, we Fourier transform the scattering matrix element in Eq.~\eqref{eq: relation} to mixed space:
\begin{widetext}
\begin{equation}
\begin{aligned}
\label{eq: scattering}
\vphantom{\int}    &\langle q(\kq',k'^+_0,\beta'_0)\Bar{q}(\kaq',k'^+_1,\beta'_1)|\hat{S}-\mathds{1}|q(\kq,k_0^+,\alpha_0)\Bar{q}(\kaq,k_1^+,\alpha_1)\rangle \\
    &\hspace{1cm}=\int_{\xq',\xaq',\xq,\xaq}e^{i\kq\xq+i\kaq\xaq-i\kq'\xq'+i\kaq'\xaq'} \delta_{\beta_0\beta_1}
    \sum_{q_0\Bar{q}_1}\Big[ U_F(\xq)_{\alpha_0\beta_0}U^\dagger_F(\xaq)_{\alpha_1\beta_1} - \delta_{\beta_0\alpha_0}\delta_{\beta_1\alpha_1}\Big] \\
    &\hspace{1.5cm}\cross\Psi(\xq,\xaq,z) \big \langle q(\xq',k'^+_0,\beta'_0)\Bar{q}(\xaq',k'^+_1,\beta'_1)|q(\xq,k_0^+,\beta_0)\Bar{q}(\xaq,k_1^+,\beta_1)\big\rangle \;. \vphantom{\int}
\end{aligned} 
\end{equation}
\end{widetext}
In the above expression, we have used the short-hand notation in Eq.~\eqref{eq: sum} for the phase space integration in momentum space as well as $\int \der^2\boldsymbol{x}_\perp=\int_{\boldsymbol{x}_\perp}$ for the Fourier transform integrals. We also omit helicity indices for simplicity, but they are easy to restore when needed since helicity is conserved in the eikonal interaction with the color field. The $U_F(\xq)$ and $U_F^\dagger(\xaq)$ Wilson lines are the result of applying the scattering operator $\hat{S}$ to the incoming quark-antiquark Fock state. It is worth noticing that with the momentum integral in the phase space (\ref{eq: sum}) and the Fourier transform exponentials, the wave function $\Psi^{\gamma\rightarrow q_0\Bar{q}_1}(\xq,\xaq,z)$ is in mixed space.

After some algebra, first using the normalization of the Fock states in Eq.~\eqref{eq: normalization}, secondly integrating over longitudinal momentum and transverse coordinates with the delta functions and finally summing over the $\beta_0,\beta_1$ color indices, we can write the scattering amplitude $\mathcal{M}$ in relation~\eqref{eq: relation} as
\begin{multline}
\label{eq: amplitude}
    \mathcal{M}_{\gamma\rightarrow q\Bar{q}}=\int_{\xq',\xaq'}  e^{i\kq\xq'}e^{i\kaq\xaq'}\psi(\xq',\xaq',z)\\
    \cross \bigg(\Big[ U_F(\xq')U^\dagger_F(\xaq')\Big]_{\beta'_0\beta'_1}- \delta_{\beta'_0\beta'_1}\bigg) \;.
\end{multline}

To calculate the differential cross section, we square the amplitude~\eqref{eq: amplitude} and sum over the color of the final state quarks:
\begin{multline}
\label{eq: square amplitude}
    \big\langle|\mathcal{M}|^2\big\rangle= \!\!\!  \int\limits_{ \substack{\xq,\xaq \\ \xq',\xaq'}} \!\!\!  \!\! e^{i\kq(\xq-\xq')}e^{i\kaq(\xaq-\xaq')}
    \big|\psi^{\gamma\rightarrow q_0\Bar{q}_1}\big|^2
\\
\times    \Big(Q_{\xq,\xaq,\xq',\xaq'}-D_{\xq,\xaq} -D_{\xaq',\xq'}+1\Big) \;.
\end{multline}
In the above expression, we introduced the dipoles $(D)$
and quadrupoles $(Q)$ which are averaged over the color configurations of the target: 
\begin{align}
    &D= \frac{1}{N_c}\big\langle\mathrm{Tr}\big[U_F(\xq)U^\dagger_F(\xaq)\big] \big\rangle_{x}\;, \label{eq: dipole}\\
    &Q= \frac{1}{N_c}\big\langle \mathrm{Tr}\big[ U_F(\xq)U^\dagger_F(\xaq) U_F(\xaq')U^\dagger_F(\xq')\big]\big\rangle_{x} \;.
\end{align}
The energy (Bjorken-$x$) dependence of dipole in Eq.~\eqref{eq: dipole} can be solved with the running coupling Balitsky-Kovchegov (rcBK) equation~\cite{Balitsky:1995ub,Kovchegov:1999yj,Balitsky:2006wa} in the limit of a large number of colors $(N_c\gg 1)$.
The $x$ dependence of $D$ and $Q$ is left implicit.
Given an initial condition at $x=x_0$, the BK equation predicts the scattering amplitude \eqref{eq: dipole} at smaller values of Bjorken-$x$. For a particle produced with momentum $k_\perp$ and mass $m$, the Bjorken-$x$ is defined as
\begin{equation}
\label{eq: bjorken x}
    x=\frac{\sqrt{m^2+k_\perp^2}}{\sqrt{s}}e^{-y} \;.
\end{equation}

Adding a phase space space integration to the amplitude square in Eq.~(\ref{eq: square amplitude}), averaging over the photon polarization and summing over the quark and antiquark helicities, we recover the well-known lowest-order differential cross section for dijet production in the CGC theory, as written, e.g., in Eq.~(22) in Ref.~\cite{Dominguez:2011wm}.    

However, for $\dm=c\Bar{u}$ photoproduction we further integrate out the antiquark momentum $\kaq$ in the differential cross section since it is not observed in the final state. While we omit the explicit calculation, we highlight that integrating over $\kaq$ in Eq.~\eqref{eq: square amplitude} -- through the associated exponential function -- yields a delta function $\delta^2(\xaq - \xaq')$. This allows for a straightforward integration over $\xaq'$. Furthermore, since the Wilson lines are unitary $(U^\dagger(\xt) U(\xt)=\mathds{1})$, the quadrupole $Q$ in Eq.~(\ref{eq: square amplitude}) simplifies to a dipole $D$:
\begin{equation}
\begin{aligned}
    &\int \der^2\xaq'\delta(\xaq-\xaq') Q (\xq,\xaq,\xaq',\xq')\\
    &=\frac{1}{\Nc} \left\langle \mathrm{Tr}\Big[ U_F(\xq)U^\dagger_F(\xaq) U_F(\xaq)U^\dagger_F(\xq')\Big] \right\rangle_x \\
    &=\frac{1}{\Nc} \mathrm{Tr} \left\langle \Big[ U_F(\xq)U^\dagger_F(\xq')\Big] \right\rangle_x  = D(\xq,\xq')\;.
\end{aligned}
\end{equation}
With this simplification our intermediate expression for charm quark production with transverse momentum $\kq$ is 
\begin{equation}
\label{eq: intermediate result}
\begin{aligned}
    &\frac{\der\sigma^{\gamma+A \rightarrow c+X}}{\der^2\kq}\\
    &\vspace{0.3cm}=\frac{\Nc}{(2\pi)^2}\int\frac{\der k_1^+}{4\pi k_1^+}\frac{\der k_0^+}{4\pi k_0^+} 2q^+(2\pi)\delta(q^+ - k_1^+ - k_0^+)\\
    &\hspace{0.5cm}\cross \int_{\xq,\xaq,\xq'}e^{i\kq(\xq-\xq')}\big|\psi^{\gamma\rightarrow q_0\Bar{q}_1}\big|^2 \\ 
    &\hspace{0.5cm}\cross\Big[D_{\xq-\xq'}-D_{\xq-\xaq}-D_{\xq'-\xaq}+1\Big] \;.
\end{aligned}
\end{equation}

Subsequently, we integrate out the longitudinal momentum $k_1^+$ of the antiquark and express Eq.~\eqref{eq: intermediate result} in terms of the longitudinal momentum fraction $z=k_0^+/q^+$ and the rapidity $y_0=\mathrm{ln}\big(\sqrt{2}k_0^+/\sqrt{(m^2+\kq^2)}\big)$ of the produced charm quark. In addition, we Fourier transform Eq.~(\ref{eq: intermediate result}) back to momentum space, which allows us to factorize the integrand into a single dipole $D(\lt)$ and a combination of momentum space  reduced wave functions $\psi$ from Eq.~(\ref{eq: LCWF 2}).

After some algebra, the differential cross section for charm quark production with momentum $\kq$ and rapidity $y_0$ from a transversely polarized photon is
\begin{widetext}
\begin{multline}
\label{eq: partonic cross section}
     \frac{\der\sigma^{\gamma_T +A \rightarrow c +X}}{\der^2\kq \dd y_0}=\frac{\aem e_q^2N_\mathrm{c}}{(2\pi)^2}\int \der^2\bt\int \frac{\der^2\lt}{(2\pi)^2}D(\ell_\perp)
     \Bigg\{2z m^2\Bigg[\frac{1}{m^2+(\kq-\lt)^2} -\frac{1}{m^2+\kq^2}\Bigg]^2  \\
     +2z\Big(z^2+(1-z)^2\Big)\Bigg[\frac{\kq-\lt}{m^2+(\kq-\lt)^2} -\frac{\kq}{m^2+\kq^2} \Bigg]^2  \Bigg\} \;.
\end{multline}
\end{widetext}
Here $\aem=e^2/(4\pi)$ is the electromagnetic coupling constant and $\Nc=3$ is the number of colors. The impact parameter $\bt$ represents the transverse distance between the photon and the target, as depicted in Fig.~\ref{fig: UPC}. The transverse momentum $\lt$ is transferred from the target to the quark-antiquark system during the interaction. In the numerical analysis presented in Sec.~\ref{sec:results}, we approximate the $\dm$ meson mass as the charm quark mass, setting $m_{\dme}\sim m=1.5~\mathrm{GeV}$.

\subsection{Inclusive \texorpdfstring{$\dm$}{\dm} photoproduction in the CGC formalism}
\label{sub: final result}
The fragmentation function $M(z_f,m)$ in Eq.~\eqref{eq: convolution} represents the hadronization of the charm quark into a $\dm$ meson, with $z_f=\kd/\kq$. We use the following lowest order parametrization from Ref.~\cite{Kniehl:2006mw}, where 
$m=1.5$ GeV, $N=0.694$ and $\epsilon=0.101$:
\begin{equation}
\label{eq: FF}
    M(z_f,m) = N \frac{z_f(1-z_f)^2}{[(1-z_f)^2+\epsilon z_f]^2} \;.
\end{equation}
We choose to evaluate the fragmentation function at the initial scale corresponding to the mass of the produced quark.
For simplicity, we also approximate the rapidity of the $\dm$ meson to be equal to the rapidity of the intermediate charm quark: $y_{0}=y_{\dme}$. 

Additionally, in our numerical analysis in Sec.~\ref{sec:results} we will also present the bottom quark contribution to $\dm$ production using the lowest-order parametrization for the $b\to \dm$ fragmentation function from Ref.~\cite{Kniehl:2006mw}.

By combining the photon flux in Eqs.~\eqref{eq:photon_flux} and~(\ref{eq:0n_emission}) with the partonic cross section~(\ref{eq: partonic cross section}) and the fragmentation function~(\ref{eq: FF}), the $\dm$ photoproduction cross section in the CGC formalism is
\begin{widetext}
\begin{equation}
\begin{aligned}
\label{eq: result cross section}
     &\frac{\der\sigma^{A+A \rightarrow \dm+X}}{\der^2\kd \der y_\mathrm{D}} \\
     &=
     \int_{0}^1\frac{\der z_f}{z_f^2}M(z_f,m)
     \int_{\frac{m_T}{\sqrt{2}}e^{y_\mathrm{D}}}^{\infty}\frac{\der q^+}{q^+}\int_{2R}^{\infty} \der B_\perp (2\pi)B_\perp P(B_\perp)F(q^+/\sqrt{2},B_\perp)\frac{\alpha_{\mathrm{e.m.}}e_f^2N_c}{(2\pi)^2}\int \der^2\bt \int \frac{\der^2\lt}{(2\pi)^2}D(\ell_\perp)\\
     & \cross 
    \Bigg\{2zm^2\Bigg[\frac{1}{m^2+(\frac{\kd}{z_f}-\lt)^2} -\frac{1}{m^2+(\frac{\kd}{z_f})^2}\Bigg]^2  +2z\Big(z^2+(1-z)^2\Big)\Bigg[\frac{\frac{\kd}{z_f}-\lt}{m^2+(\frac{\kd}{z_f}-\lt)^2} -\frac{\frac{\kd}{z_f}}{m^2+(\frac{\kd}{z_f})^2} \Bigg]^2  \Bigg\} \;.
\end{aligned}
\end{equation}
\end{widetext}
In the above expression, we have related the photon energy $\omega$ in Eq.~\eqref{eq:photon_flux} to the photon plus momentum as
\begin{equation}
    q^+=\frac{1}{\sqrt{2}}(\omega + q^3)=\sqrt{2}\omega \;.
\end{equation}
Since the longitudinal momentum fraction $z=k_0^+/q^+$ in the $\gamma \to c\bar{c}$ splitting is constrained in $0<z<1$, for a fixed meson momentum and fragmentation $z_f$ this imposes a lower limit on $q^+$: $m_T/\sqrt{2}e^{y_\mathrm{D}}=k_0^+ <q^+$, where the quark transverse mass is $m_T=\sqrt{m^2+(\kd/z_f)^2}$. Similarly, the integral over $z_f$ is also constrained in $0<z_f<1$.

\subsection{The collinear factorization limit}
\label{sub: collinear approximation}
Before moving to our numerical analysis, let us first analyze our result \eqref{eq: result cross section} in more detail. In particular we want to demonstrate how to take the collinear limit of our calculation. We will see that the cross section reduces to what can clearly be identified as the hard factor and the collinear gluon PDF as they would appear in the  corresponding collinearly factorized expression. 

In the actual calculation we will take the dipole operator from a parametrization satisfying the Balitsky-Kovchegov evolution equation~\cite{Balitsky:1995ub,Kovchegov:1999yj} with initial conditions fit to HERA data. However, for the purposes of this discussion it is better to use a more tractable analytical parametrization. For this purpose we will here take the dipole in Eq.~(\ref{eq: dipole}) to be given by the explicit solution in the McLerran-Venugopalan (MV) model \cite{McLerran:1993ka,McLerran:1993ni,McLerran:1994vd}: 
\begin{equation}
\label{eq: dipole in MV}
   D_{\mathrm{MV}}(r=|\xt-\yt|)=\mathrm{exp}\Bigg(-\frac{\Qs^2 r^2}{4} \mathrm{ln}\frac{1}{r\Lambda} \Bigg) \;,
\end{equation}
where $\Qs$ is known as the saturation scale and the parameter $\Lambda$ is an infrared cutoff.

The transverse momentum dependence of the cross section originates from  the $\lt$--integral in Eq.~\eqref{eq: result cross section}:
\begin{equation}
\label{eq: ell integral}
     \mathrm{I}_{\ell}=\int \frac{\der^2\lt}{(2\pi)^2}D(\ell_\perp)\{\dots\} \;.
\end{equation}
The dipole $D(\ell_\perp)$ in momentum space (Fourier transform \eqref{eq: Fourier transform} of Eq.~\eqref{eq: dipole in MV}) follows the approximate scaling behavior:
\begin{equation}
\label{eq: dipole behavior}
    D(\ell_\perp)\sim
    \begin{cases}
     \mathrm{const.},& \ell_\perp \ll \Qs \: \mathrm{(saturated\;regime)}\\
     1/\ell_\perp^4, & \ell_\perp \gg \Qs \:\mathrm{(dilute\;regime)}\;.
    \end{cases}
\end{equation}
The saturation scale $\Qs$ separates the perturbative QCD regime, governed by linear evolution equations (DGLAP and BFKL), from the non-linear saturation regime described by the BK equation~\cite{Kovchegov:2012mbw}. 
For charm quarks at LHC energies, in practice\footnote{We refer the reader to Ref.~\cite{Lappi:2013zma} for the definition of the saturation scale $\Qs$ used in this work.} $\Qs \sim m \approx 1.5$~GeV, making charm observables sensitive probes of gluon saturation.

The dipole amplitude is multiplied  in Eq.~\eqref{eq: result cross section} by  the expression inside the braces, which is a kind of ``hard factor'' for charm production. In both terms of this expression the term inside the
square brackets $[\dots]$ vanishes linearly in  $\ltm$ in the limit $\ell_\perp\rightarrow0$. On the other hand, for $\ell_\perp \gg  k_{0\perp} = k_{D\perp}/z_f$ it approaches a constant. This means that the term inside the braces follows the approximate scaling: 
\begin{equation} \label{eq:hardfactbehavior}
    \{\dots\}\sim
    \begin{cases}
     \ell_\perp^2,& \ell_\perp \ll k_{0\perp} \;, \\
     \mathrm{const.}, & \ell_\perp\gg k_{0\perp} \;. 
    \end{cases}
\end{equation}

We expect that for large momenta, $\kqm\gg \Qs$, the CGC expressions should exhibit a behavior corresponding to a collinear factorization picture of the same process. Let us therefore  analyze how the partonic cross section in Eq.~\eqref{eq: partonic cross section} behaves in the limit $k_{0\perp}\gg \Qs$.
Combining the estimates \eqref{eq: dipole behavior}  and~\eqref{eq:hardfactbehavior} we see that the integral $\mathrm{I}_{\ell_\perp}$  in Eq.~\eqref{eq: ell integral} exhibits a different behavior in three distinct kinematic regimes: 
\begin{align}
 &\mathrm{(A)} \ \ \ell_\perp \ll \Qs \ll \kqm: \int^{\Qs}_0 \der\ell_\perp \ell_\perp^3 \\
 &\mathrm{(B)} \ \ \Qs \ll \ell_\perp \ll \kqm: \int^{\kqm}_{\Qs}\der\ell_\perp \frac{1}{\ell_\perp} \label{eq: dilute region}\\
 &\mathrm{(C)} \ \ \Qs \ll \kqm \ll \ell_\perp: \int_{\kqm}^\infty\der\ell_\perp \frac{1}{\ell_\perp^3} \label{eq: power tail}\;.
\end{align} 
We observe that the integral from region (B) is logarithmic. This logarithm is the dominant contribution in the regime of validity of collinear factorization $\kqm\gg \Qs$. 
The usual way of digging this limit out of the full expression~\eqref{eq: partonic cross section} is the following. We first start from saying that typically $\ltm \sim \Qs \ll \kqm$, and develop the term inside braces in the limit $\ell_\perp \ll \kqm$  to get  
\begin{multline}
\label{eq: dilute limit cross section}
     \mathrm{I}_{\ell_\perp}^{\mathrm{collinear}}=\int \frac{\der^2\lt}{(2\pi)^2}D(\ell_\perp)\lt^2\\
     \cross\Bigg\{2z\frac{2 m^2k_{0\perp}^2 + [z^2+(1-z)^2](k_{0\perp}^4+m^4)}{(k_{0\perp}^2+m^2)^4} \Bigg\}
     \;.
\end{multline}
Now, in principle, the upper integration limit of $\ell_\perp$ is infinity, and with the  behavior~\eqref{eq: dipole behavior}  the integral does not actually converge. However, this divergence is just an artefact from incorrectly using the $\ell_\perp \ll k_{0\perp}$ behavior of the hard factor in regime C, which does not give a logarithmic contribution. It can therefore be fixed by setting an upper limit $\ell_\perp\leq k_{0\perp}$ or $\ell_\perp\leq k_{D\perp}$ (the choice of which one does not affect the leading logarithm). This turns the cross section into a product of a hard factor which is independent of $\ell_\perp$, and an integrated gluon distribution 
\begin{equation}\label{eq:defxg}
xg(x,k_{0\perp}^2) \propto    \int^{k_{0\perp}} \frac{\der^2\lt}{(2\pi)^2}D(\ell_\perp)\lt^2 \;,
\end{equation}
which depends on the hard scale of the process $k_{0\perp}^2$. With the $D(\ell_\perp)\sim 1/\ell_\perp^4$ behavior in the MV model~\eqref{eq: dipole in MV}, the integrated gluon distribution behaves logarithmically in the hard scale, which corresponds to the DGLAP evolution of the gluon distribution. 
This procedure can also be thought of as the limit  being  $\ell_\perp \ll m$, if we assume the quark mass to be the large scale, and the collinear factorization regime as $m\gg \Qs$.

In the next section, we will numerically compare the full integral $\mathrm{I}_{\ell_\perp}$ in Eq.~\eqref{eq: ell integral} to the collinear approximation $\mathrm{I}_{\ell_\perp}^{\mathrm{coll.}}$ in Eq.~\eqref{eq: dilute limit cross section} for a fixed photon energy $\omega$. 
We believe this is a more informative apples-to-apples comparison of the CGC formalism and collinear factorization than a full calculation with state-of-the art DGLAP evolve parton distribution. By comparing the CGC calculation with its own collinear limit, we can better assess the effect of the $\Qs$--scale transverse momentum $\ell_\perp$ on observables. The comparison we make in the next section focuses on the transition region from the saturation regime $\kqm\sim \Qs$ to the collinear regime $\kqm\gtrsim \Qs$, which is more interesting for saturation physics. This differs from the correct resummation of the logarithms $\as \ln k_{0\perp}^2/\Qs^2$, which  is, at least parametrically, only relevant for larger values of $k_{0\perp}$, and can only be performed correctly by a proper DGLAP evolution of the integrated gluon distribution.

\section{Numerical Predictions}
\label{sec:results}
Given the differential cross section in Eq.~\eqref{eq: result cross section}, we next make predictions for inclusive $\dm$ photoproduction in lead-lead UPCs at the center-of-mass collision energy $\sqrt{s}=5360\:\mathrm{GeV}$. Following Ref.~\cite{Lappi:2013zma}, we evolve the dipole correlator in Eq.~\eqref{eq: result cross section} with the Balitsky-Kovchegov (rcBK) equation~\cite{Balitsky:1995ub,Kovchegov:1999yj,Balitsky:2006wa} with an initial condition parametrized at $x_0=0.01$. This initial condition for the dipole-proton scattering has a form inspired by the McLerran-Venugopalan (MV) model as~\cite{McLerran:1993ni} 
\begin{equation}
\label{eq: initial condition dipole}
    \mathcal{N}(\boldsymbol{r}_\perp)=1-\mathrm{exp} \Bigg[-\frac{(\boldsymbol{r}_\perp^2 Q^2_{s0})^\gamma}{4}\mathrm{ln}\Bigg(\frac{1}{|\boldsymbol{r}_\perp|\Lambda_\mathrm{QCD}}+e_c\cdot e \Bigg) \Bigg] \;,
\end{equation}
where $\mathcal{N}(\rt)=1-D(\rt)$. Here $\Lambda_\mathrm{QCD}=0.241$ GeV and we use the MV$^e$ fit of Ref.~\cite{Lappi:2013zma}, where  $\gamma$ is fixed to  $\gamma=1$ and the free parameters $Q_{s0}^2=0.060~\mathrm{GeV^2}$ and $e_c=18.9$ are determined by a fit to the HERA inclusive DIS cross section data~\cite{H1:2009pze}. For a nuclear target, the rcBK evolution equation  is solved separately for each impact parameter $b_\perp$ using an initial condition where the $b_\perp$--dependence of the saturation scale is obtained from the optical Glauber model:
\begin{multline}\label{eq:Aic}
    \mathcal{N}^A(\boldsymbol{r}_\perp,\boldsymbol{b}_\perp)=1-\mathrm{exp}{\Bigg[ -AT_A(\boldsymbol{b}_\perp) \pi R_p^2 \frac{(\boldsymbol{r}_\perp^2Q_{s0}^2)^\gamma}{4} \Bigg]}\\
    \cross \mathrm{ln}\Bigg(\frac{1}{|\boldsymbol{r}_\perp|\Lambda_{QCD}}+e_c\cdot e \Bigg) \;.
\end{multline}
Here $T_A(\boldsymbol{b}_\perp)$ is the Woods-Saxon nuclear density normalized to unity: $\int \dd[2]{\bt} T_A(\bt)=1$. In the region where the nuclear saturation scale would fall below that of the proton, we replace the dipole-nucleus scattering amplitude by the dipole-proton amplitude scaled such that all non-trivial nuclear effects vanish. Further details can be found in Ref.~\cite{Lappi:2013zma}.
We emphasize that in Eq.~\eqref{eq:Aic} there are no additional free parameters for the nucleus apart from the standard Woods-Saxon density. Thus, the nuclear modification of cross sections is a pure prediction of the framework without any additional fit parameters.

In Table~\ref{tab:Bjorken} we present the Bjorken-$x$ values reached in lead-lead UPCs in CMS kinematics covering rapidities $-2<y_\mathrm{D}<2$ at the center-of-mass energy of $\sqrt{s}=5.36\;\mathrm{TeV}$. 
The CMS data is found to probe the lead nucleus from the initial condition of the small-$x$ evolution ($x=x_0=0.01$) down to very small $x=3\cdot 10^{-5}$, which makes this data a potentially powerful probe of the small-$x$ evolution.
\begin{table}[tb]
    \centering
    \renewcommand{\arraystretch}{1.1} 
    \begin{tabular}{c @{\hspace{0.5cm}} c @{\hspace{0.5cm}}c}  
        \toprule
        & \multicolumn{2}{c}{$x$} \\  
        \cmidrule(lr){2-3}  
        $y_\mathrm{D}$ & $k_{\mathrm{D}\perp} = 0$ GeV & $12$ GeV \\  
        \midrule
        $-2$ & $2 \cdot 10^{-3}$ & $0.01$ \\
        $-1$ & $7\cdot 10^{-4}$ & $6\cdot 10^{-3}$ \\  
        $0$  & $2\cdot 10^{-4}$ & $2\cdot 10^{-3}$ \\
        $1$  & $1\cdot 10^{-4}$ & $8\cdot 10^{-4}$ \\
        $2$  & $3\cdot 10^{-5}$ & $3\cdot 10^{-4}$ \\
        \bottomrule
    \end{tabular}
    \caption{Bjorken-$x$ values for fixed rapidity $y_\mathrm{D}$ and transverse momentum $k_{\mathrm{D}\perp}$ of the produced $\dm$ meson. We use $m=1.5\;\mathrm{GeV}$ and $\sqrt{s}=5360\;\mathrm{GeV}$ in Eq.~\eqref{eq: bjorken x}. For rapidity $y_\mathrm{D} = -2$, we set a maximum value of $x=0.01$ in the momentum range $7 <\kdm\leq 12 \;\mathrm{GeV}$.}
    \label{tab:Bjorken}
\end{table}

\begin{figure}[tb]
    \centering
    \includegraphics[width=1\linewidth]{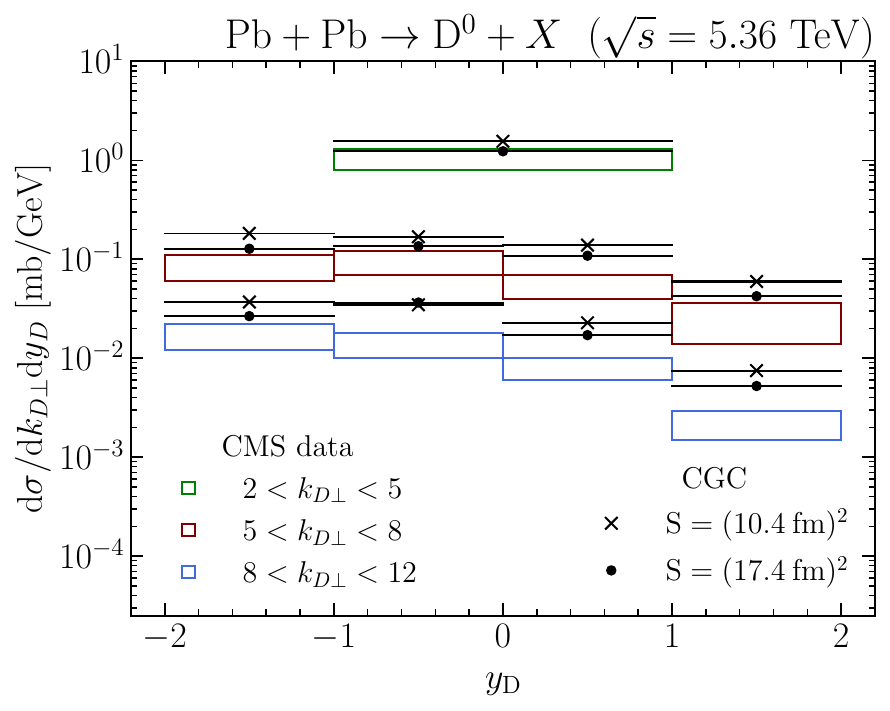}
    \caption{Differential prompt $\dm$ photoproduction cross section in UPC as a function of the $y_\mathrm{D}$ rapidity of the produced meson. The CGC theory predictions using the two parameters  $S$ in the photon flux are integrated over the same momentum and rapidity bins as the preliminary CMS inclusive data in Ref.~\cite{CMS}, which is represented by the color boxes. }
    \label{fig: CMS comparison}
\end{figure}

We begin the numerical analysis by quantifying the bottom  quark contribution to prompt $\dm$ photoproduction in CMS kinematics. The relative importance of the $b$ channel is shown in Table~\ref{tab:sigma_ratio}. 
This contribution is overall small, maximally $\sim 10\dots 15\%$.
Looking first at the rapidity dependence for a fixed $k_{\mathrm{D}\perp}$, the decrease of the bottom contribution with increasing rapidity is related to the larger mass of the bottom quark. Due to the mass difference, the $b$ quark production requires higher photon energies $\omega$ which are strongly suppressed by the Bessel function in the photon flux, see Eq.~\eqref{eq:photon_flux}. The photon flux is also the reason why  the photoproduction cross section decreases rapidly toward larger rapidities (see Fig.~\ref{fig: CMS comparison}).

On the other hand, the transverse momentum dependence in Table~\ref{tab:sigma_ratio} 
can be explained with the integrand in Eq.~\eqref{eq: result cross section}. This roughly behaves as $1/((\frac{\kd}{z_f})^2+m^2)^n$, where $n$ varies due to the lower limit of the photon energy integral. If we take the limits $\kdm\rightarrow 0$ and $\kdm\rightarrow \infty$ to the integrand, we get: 
\begin{equation}
\begin{aligned}
    &\frac{\sigma_b}{\sigma_c}\sim\bigg(\frac{m_c}{m_b}\bigg)^n, & & \kdm\rightarrow 0 \;, \\
    &\frac{\sigma_b}{\sigma_c}\rightarrow 1 \;, & &\kdm\rightarrow \infty \;.
\end{aligned}
\end{equation}
Therefore, in the small momentum region, $0<k_{\mathrm{D}\perp}\lesssim5\,\mathrm{GeV}$, the mass of the quarks is relevant and the bottom contribution is small. In contrast, at large $k_{\mathrm{D}\perp}$ both quarks behave as ``light'', which results in an increase of the bottom to $\dm$ contribution. 

The CGC framework describes nonlinear gluon saturation effects at $\kdm^2\sim Q_s^2$. In this region, the transverse momentum of the produced quark pair originates not only from the back-to-back recoil of the charm quarks but also from the classical gluon field in the nucleus. 
As such, we are mostly interested in $\dm$ production at low $\kdm \lesssim 5\,\mathrm{GeV}$ where saturation effects are expected to be largest. In this kinematical domain the $b$ quark contribution is less than $5\%$ and it is neglected from now on. 
\setlength{\tabcolsep}{5pt} 
\renewcommand{\arraystretch}{1.2} 
\begin{table}[tb]
    \centering
    \begin{tabular}{c c c c c}
        \toprule
        & \multicolumn{4}{c}{$\sigma_{b\rightarrow \dm}/\sigma_{c\rightarrow \dm}$} \\  
        \cmidrule(lr){2-5}
         & \centering {$y_\mathrm{D}=0$}  & {$0.5$} &{$1$} & {$1.5$} \\  
        \midrule
        $0<k_{\mathrm{D}\perp}<5 $  & 5-5\% & 5-5\%  & 4-4\%  & 3-3\%  \\  
        $5<k_{\mathrm{D}\perp}<12$ & 14-13\% & 13-12\% &  10-10\% &8-7\% \\  
        \bottomrule
    \end{tabular}
        \caption{Ratio of bottom over charm contribution to $\dm$ production integrated over the $k_{\mathrm{D}\perp}$ momentum range (GeV) and fixed $y_\mathrm{D}$ rapidity values of the meson. The first value corresponds to $S=(10.4\;\mathrm{fm})^2$ in Eq.~\eqref{eq:0n_emission} and the second to $S=(17.4\;\mathrm{fm})^2$. 
        }
        \label{tab:sigma_ratio}
\end{table}

\begin{figure}[tb]
    \centering
    \includegraphics[width=1\linewidth]{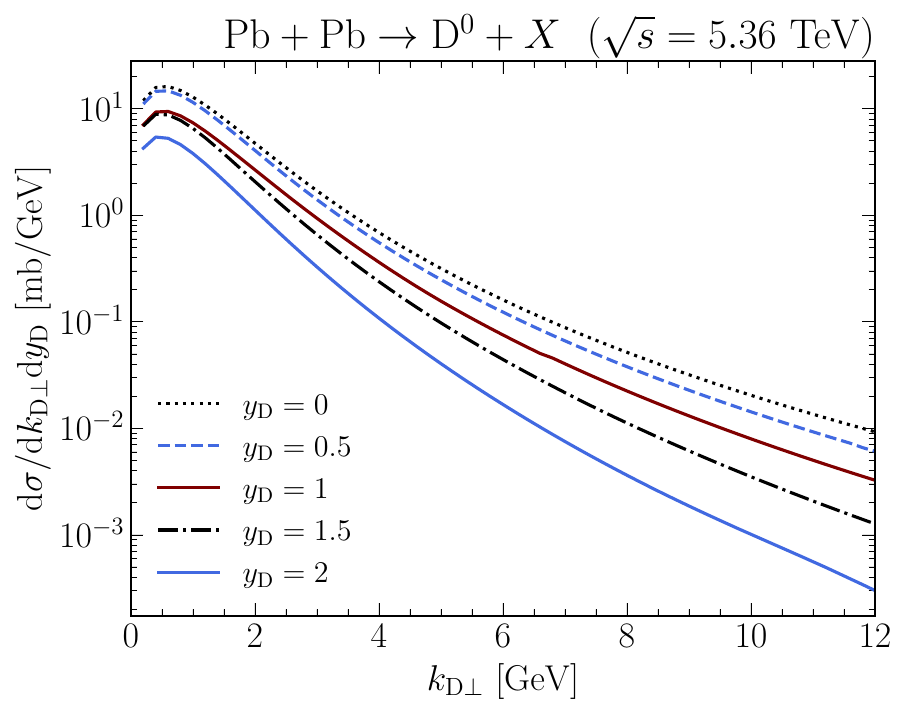}
    \caption{Differential $\dm$ photoproduction cross section in UPC as a function of the transverse momentum $\kdm$ of the produced meson and fixed rapidities. Here we use  $S=(17.4 \; \mathrm{fm})^2$ in Eq.~\eqref{eq:0n_emission} in order to enforce the  ``0n'' neutron multiplicity class.}
    \label{fig: cross section}
\end{figure}

The transverse momentum distribution of prompt $\dm$ photoproduction in UPCs is 
presented in Fig.~\ref{fig: cross section}.  A comparison to  the preliminary CMS data~\cite{CMS} is shown in Fig.~\ref{fig: CMS comparison}. We find that the data is well described in the small-$\kdm$ region, $2<\kdm<5\;\mathrm{GeV}$, specially for $S=(17.4\;\mathrm{fm})^2$. This result aligns with the CGC framework applied. Our predictions overestimate the data at higher $\kdm$, meaning that our cross section does not decrease with $\kdm$ as steeply as the measurement. We also observe that a larger value of $S=(17.4\ \mathrm{fm})^2$ in the electromagnetic dissociation factor in Eq.~\eqref{eq:0n_emission}, which favors smaller $B_\perp$ values, leads to a stronger rapidity dependence  and a better agreement with the data. It would be very interesting to see experimental data with a finer binning in $\kdm$. 

When comparing our results in Fig.~\ref{fig: cross section} with previous predictions for open charm photoproduction, e.g. \cite{Klein:2002wm, Goncalves:2017zdx}, we observe that the $\kdm$ spectra have a similar shape and peak position, although our predicted maximum value is smaller. At present, we do not have an explanation for this difference.

It is worth noting that the parameters of the initial condition in Eq.~\eqref{eq: initial condition dipole}, along with the proton transverse size $\sigma_0/2=16.36\;\mathrm{mb}$ used in this work, were obtained by fitting the HERA proton structure function data in Ref.~\cite{Lappi:2013zma} considering only light quarks.
If a similar fit were done including heavy quarks, a smaller saturation scale and/or proton size would be obtained, resulting in smaller $\dm$ photoproduction cross section. However, no leading order fit that simultaneously describes the total cross section and charm production in DIS exists\footnote{We note that in Ref.~\cite{Albacete:2010sy} both the total cross section and charm production data are simultaneously described, with the caveat that one has to introduce a separate parametrization for the $c\bar c$-target scattering.} (unlike at NLO~\cite{Hanninen:2022gje}), and as such we adopt the MV$^e$ fit from~\cite{Lappi:2013zma}.

Due to this uncertainty in the heavy quark dipole-target scattering amplitude, we can not rigorously estimate the uncertainty in our predictions. As such, we also do not try to estimate uncertainties from the non-perturbative fragmentation functions (e.g. in terms of scale variations). However, 
recently, uncertainty estimates for the dipole amplitude (in the same three flavor scheme) have been obtained in Ref.~\cite{Casuga:2023dcf}. In Appendix~\ref{sec: BK uncertainty} we show that the uncertainty in the BK initial condition (without heavy quarks) has a few percent effect on the $\dm$ photoproduction cross section in LHC kinematics. 

We also quantify saturation effects in terms of the nuclear modification factor $R_{pA}$. This is defined as 
\begin{equation}
    \label{eq: RpA}
    R_{pA}=\frac{\frac{\der \sigma_{AA}}{\der^2\kd\der y_{\mathrm{D}}}}{A\cross\frac{\der \sigma_{pA(B_\perp>2R)}}{\der^2\kd\der y_{\mathrm{D}}}} \;.
\end{equation}
Here, $B_\perp>2R$ refers to the fact that we calculate the baseline using the same photon flux as in $\mathrm{Pb}+\mathrm{Pb}$ collisions. 
As such, this ratio is not directly measurable, but provides a convenient method to determine the magnitude of the saturation effects. In particular, in the dilute limit (at large $\kdm$) one obtains $R_{pA}\to 1$.

\begin{figure}[tb]
    \centering
    \includegraphics[width=1\linewidth]{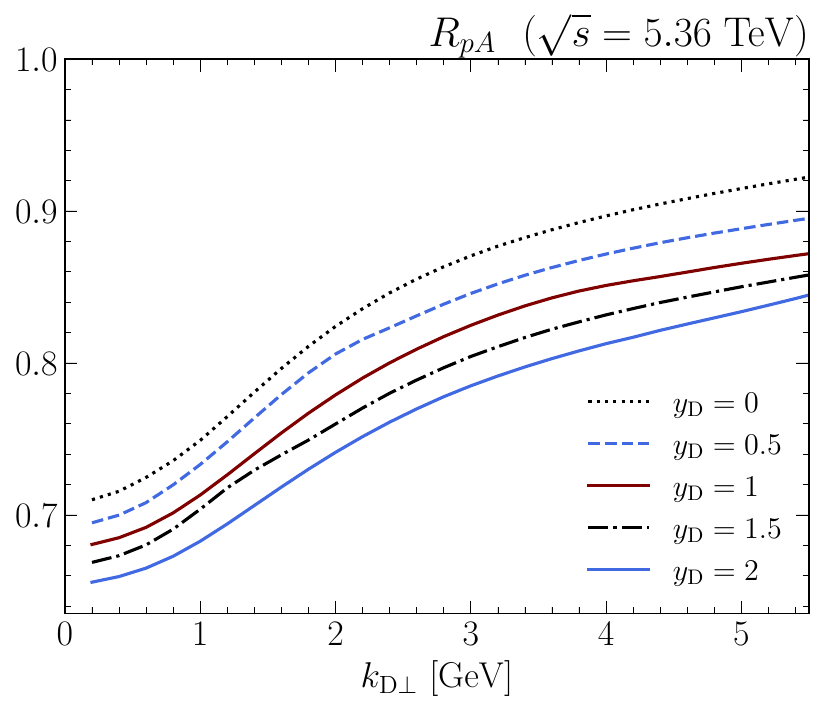}
    \caption{Nuclear modification factor $R_{pA}$ as a function of the transverse momentum $\kdm$ of the produced $\dm$ meson for fixed rapidities $y_\mathrm{D}\in[0,2]$.}
    \label{fig: RpA}
\end{figure}

The obtained nuclear modification factors at different rapidities are shown in Fig.~\ref{fig: RpA}, where we predict a strong nuclear suppression, $R_{pA}\approx 0.7$, at low $\dm$ transverse momentum. A slightly stronger suppression is obtained at more forward rapidities, and the rapidity dependence is similar to what is obtained in Refs.~\cite{Lappi:2013zma,Ducloue:2017kkq} in the case of light hadron and isolated photon production in proton-lead collisions.

\begin{figure}[tb]
    \centering
    \includegraphics[width=1\linewidth]{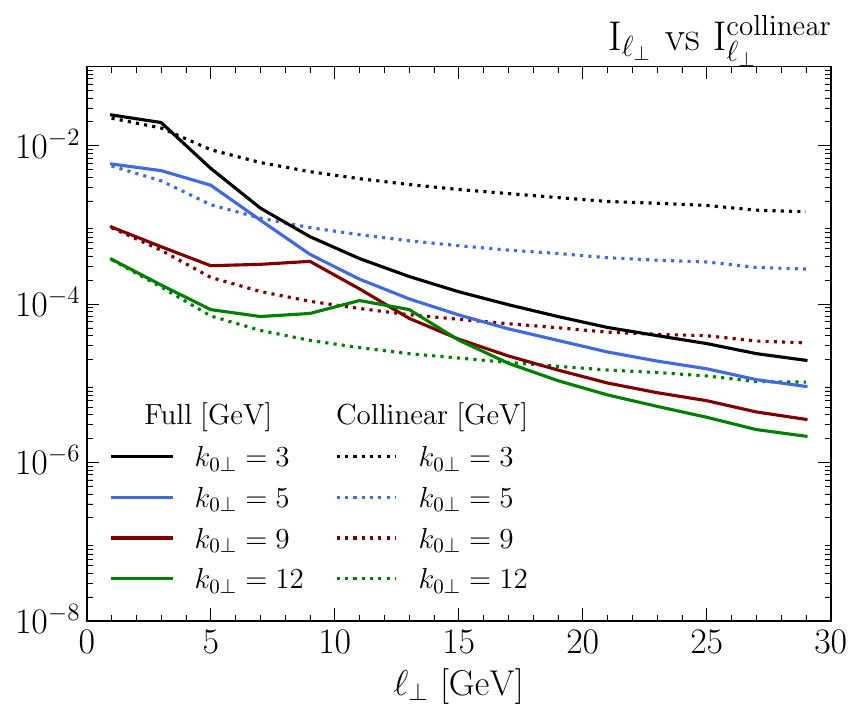}
    \caption{Open charm production at fixed momentum $k_{0\perp}$, calculated from Eq.~\eqref{eq: ell integral} (full) and Eq.~\eqref{eq: dilute limit cross section} (collinear approximation), as a function of the transverse momentum $\ell_\perp$ transferred from the target. In this photon-nucleus collision we fix the impact parameter to $b_\perp=2\;\mathrm{GeV}^{-1}$, the rapidity to $y=0$ and the photon longitudinal momentum to $q^+=50\;\mathrm{GeV}$.  }
    \label{fig: l_integral}
\end{figure}

\begin{figure}[tb]
    \centering
    \includegraphics[width=1\linewidth]{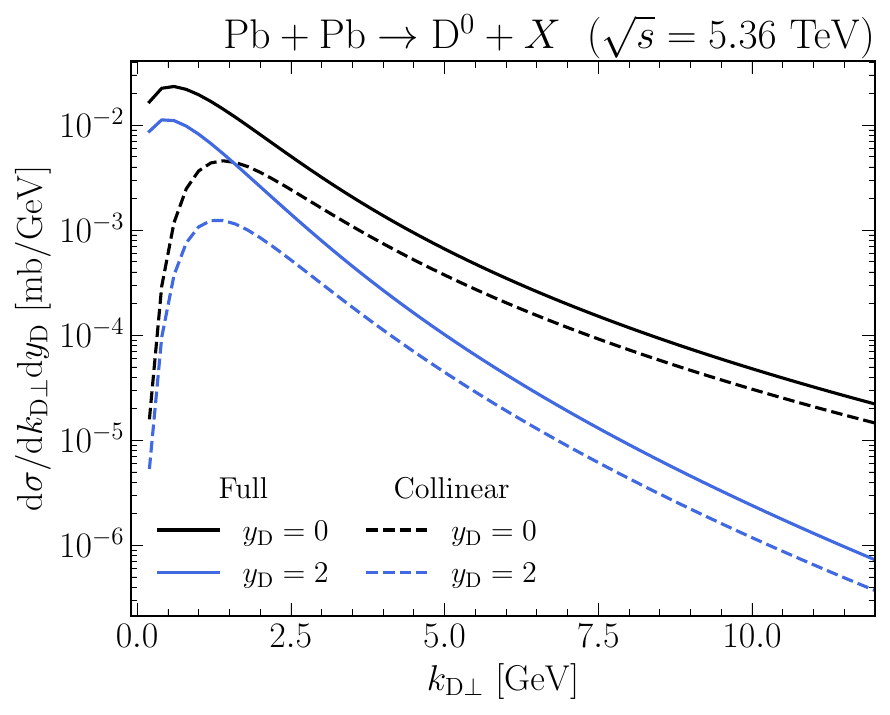}
    \caption{$\dm$ photoproduction cross section calculated from Eq.~\eqref{eq: result cross section} (full) and its collinear approximation expressed in Eq.~\eqref{eq: dilute limit cross section} (collinear) as a function of the meson transverse momentum $k_{\mathrm{D}\perp}$. For simplicity we fix the impact parameter to $b_\perp=2\;\mathrm{GeV}^{-1}$.}
    \label{fig: full_coll}
\end{figure}

To conclude the numerical analysis, we study the collinear factorization limit of the $\dm$ differential cross section in Eq.~\eqref{eq: result cross section}. Note that we are here not numerically comparing our result to a  collinear factorization result with standard PDFs, but rather to the collinear limit of our calculation.
This comparison enables us to more cleanly understand genuine saturation effects in the cross section from ones that just follow from different collinear gluon PDF sets having different numerical values.
First, in Fig.~\ref{fig: full_coll} we compare the behavior of the $\lt$--integrand for charm photoproduction in Eq.~\eqref{eq: ell integral} to its collinear approximation $(\ltm\rightarrow0)$ expressed in Eq.~\eqref{eq: dilute limit cross section}. The full result exhibits, most clearly for larger $\kqm$,  the expected behaviour where the decrease is more moderate up to $\ltm \sim \kqm$ as in Eq.~\eqref{eq: dilute region}, followed by a steeper power law for $\ltm \gg\kqm$ as in  Eq.~\eqref{eq: power tail}. For the collinear limit, the behavior corresponding to a logarithmic contribution  as in Eq.~\eqref{eq: dilute region} persists at all momenta. Notable is also the local maximum structure around $\ltm \approx \kqm$, which is easily understood from the $(\lt-\kq) ^2+m^2$ denominator and is not captured by the collinear approximation at all.  The collinear and full  expressions coincide only at large $\kqm$ and small $\ltm$, which is most clearly visible in the green lines ($\kdm=12\,\mathrm{GeV}$) in Fig.~\ref{fig: l_integral}.

Subsequently, we compare the CGC expression for $\dm$ photoproduction in Eq.~\eqref{eq: result cross section} with its collinear factorization limit as a function of $\kdm$. As explained in subsection \ref{sub: collinear approximation}, to make the logarithmic expression in Eq.~\eqref{eq: dilute limit cross section} to converge, we set the upper limit in the $\lt$--integral to $\kdm$.  As expected from Fig.~\ref{fig: l_integral}, we confirm in Fig.~\ref{fig: full_coll} that the CGC and the collinear expressions do not coincide throughout the $\kdm$ range we are considering. At small $\kdm\sim \Qs$ the collinear approximation is suppressed completely unrealistically. This is  because in this regime, $D(\ltm)$ is very far from the MV model power law, and the definition~\eqref{eq:defxg} of the integrated gluon distribution with a cutoff at the scale $\kqm$ is not very reasonable. A typical DGLAP--evolved PDF would have a smoother scale dependence in practice. However, the significant deviation observed in Fig.~\ref{fig: full_coll} shows that gluon saturation is the dominant physical effect in this region. At higher $\kdm$ the full CGC calculation and the collinear limit qualitatively agree, as expected. In the way the we have obtained the  collinear approximation here the normalization is somewhat uncontrolled, which in this case leads to a suppression by a factor 2\dots 3 compared to the full result. 

\section{Summary}
\label{sec: summary}
In this paper we have derived in Eq.~\eqref{eq: result cross section} the differential cross section for inclusive $\dm$ photoproduction in lead-lead UPCs in the CGC formalism, at lowest order in the strong coupling constant. Our intermediate result for quark pair production in Eq.~\eqref{eq: partonic cross section} agrees with Ref.~\cite{Dominguez:2011wm}, and our final expression for single inclusive charm photoproduction is consistent with Refs.~\cite{Marquet:2009ca,Goncalves:2017zdx}. 

Concretely, in this work we apply this framework in a comparison to CMS data for the ``Xn0n'' event class where the projectile nucleus remains intact after the interaction.  This constraint is imposed in the cross section by a probability distribution, Eq.~\eqref{eq:0n_emission}, for the ``0n'' emission in the projectile. In contrast, the target nucleus always dissociates in inclusive $\dm$ production, emitting ``X'' neutrons in its forward region.  We leave for a future publication a more refined calculation of this breakup survival factor along the lines in  Ref.~\cite{Eskola:2024fhf}, where it is calculated using the optical Glauber model. 
The authors of Ref.~\cite{Eskola:2024fhf} found that modifying the photon flux to account for the nuclei as extended charge distributions significantly affects the highest-energy photons $(\omega=10^{-2}\sqrt{s}/2)$. This is not significant in the kinematical domain considered in this work.

The CGC description of the dense target is constrained by HERA DIS data~\cite{Lappi:2013zma}, and the hadronization process is described with a lowest-order fragmentation function~\cite{Kniehl:2006mw}. Thus, there are no free parameters in our setup and we calculate $\dm$ photoproduction consistently with the DIS data. 
A relatively good description of the CMS data is obtained especially at low $\dm$ transverse momenta, although the data would prefer a more steeply falling spectrum.

We also compare the collinear factorization limit of our calculation with the full result in Figs.~\ref{fig: l_integral} and~\ref{fig: full_coll}. This comparison first of all  shows that there is no fundamental disagreement, the two approaches are perfectly consistent qualitatively in the region where both are applicable. At $\kdm \lesssim 2\;$GeV the transverse momentum dependence starts to be strongly affected by gluon saturation, which the collinear factorization limit does not treat correctly. On the other hand, at high $\kdm$ large logarithms of transverse momenta need to be resummed with a DGLAP evolution of the collinear gluon distribution, which we do not do here. 

We expect that including heavy quarks into the DIS fits for the initial condition parametrization of the dipole amplitude in Eq.~\eqref{eq: dipole in MV} could improve the normalization of the cross section. This would lead to a better agreement with the CMS data at large transverse momentum. Furthermore, we emphasize the lack of experimental data for $\kdm<2\;\mathrm{GeV}$, which is a region of most interest for CGC calculations where a strong saturation effect (nuclear modification factor $\sim 0.7$) is predicted.   Unfortunately, at very small $\kdm$ the collinear fragmentation approach that we use for hadronization might become more questionable. Overall, the smooth matching of CGC calculations to DGLAP evolution in terms of either integrated or transverse momentum dependent PDFs has been actively discussed recently \cite{Altinoluk:2019wyu,Boussarie:2020fpb,Boussarie:2021wkn,Hauksson:2024bvv,Caucal:2024vbv,Caucal:2024bae,Caucal:2024nsb}. We believe the inclusive hadron production in $\gamma+A$ or $\gamma+p$ collisions is an interesting process both experimentally and theoretically for understanding this physics.

\begin{acknowledgments}
We thank Farid Salazar and Gian Michele Innocenti for useful discussions. 
This work was supported by the Research Council of Finland, the Centre of Excellence in Quark Matter (projects 346324 and 364191), and projects 338263, 346567 and 359902, and by the European Research Council (ERC, grant agreements  No. ERC-2023-101123801 GlueSatLight and ERC-2018-ADG-835105 YoctoLHC).
Computing resources from CSC – IT Center for Science in Espoo, Finland and the Finnish Grid and Cloud Infrastructure (persistent identifier \texttt{urn:nbn:fi:research-infras-2016072533}) were used in this work.
The content of this article does not reflect the official opinion of the European Union and responsibility for the information and views expressed therein lies entirely with the authors. 

\end{acknowledgments}

\appendix
\section{General conventions}
\label{sec: Appendix}

The transversely polarized photon 4-vector in the light-cone gauge is
\begin{equation}
    \epsilon^\mu_{\lambda=\pm}(q)=\left(0,\frac{\qt \cdot \boldsymbol{\varepsilon}_\lambda}{q^+},\boldsymbol{\varepsilon}_\lambda\right)\;.
\end{equation}
This is orthonormal and complete:
\begin{align}
    \epsilon^\mu(q,\lambda)\epsilon^*_\mu(q,\lambda')=-\delta_{\lambda,\lambda'}\;, & & q^\mu\epsilon_\mu(q,\lambda)=0\;.
\end{align}
The circular polarization $\boldsymbol{\varepsilon}_\lambda$ can be written as
\begin{equation}
    \label{eq: Circular polarization operators}
    \boldsymbol{\varepsilon}_{\pm}=\frac{1}{\sqrt{2}} \begin{pmatrix} \mp 1 \\ -i \end{pmatrix} \:.
\end{equation}
In subsection~\ref{subsec: Light-Front quantization}, we use the following convention for the two-dimensional Fourier transforms between mixed and momentum space: 
\begin{align}
\label{eq: Fourier transform}
    f(\boldsymbol{p}_\perp)&=\int \dd[2]\xt e^{i\pt\xt}f(\xt) \;,\\
    f(\xt)&=\int \frac{\dd[2]\pt}{(2\pi)^2} e^{-i\pt\xt}f(\pt) \;.
\end{align}
In mixed space, quark eigenstates with light-front helicity $h$ and color $c$ in the fundamental representation are normalized as 
\begin{equation}
\label{eq: normalization}
\begin{aligned}
    &\big\langle q(p^+,\xt,h_p,c_p)\big|q(q^+,\yt,h_q,c_q)\big\rangle\\
    &= 2p^+(2\pi)\delta(p^+-q^+)\delta^{(2)}(\xt-\yt)\delta_{h_p,h_q}\delta_{c_p,c_q} \;.
\end{aligned}
\end{equation}
\section{Uncertainty estimate from the BK initial condition}
\label{sec: BK uncertainty}
Main results of this work have been obtained by applying the MV$^e$ fit of Ref.~\cite{Lappi:2013zma} to set the initial condition for the small-$x$ evolution of the dipole amplitude. This fit, however, does not provide any uncertianty estimates. Recently, a new fit that also provides a method to propagate uncertainties in the initial condition to other observables depending on the dipole amplitude has been published in Ref.~\cite{Casuga:2023dcf}. In this Appendix, we estimate the uncertainty of our predictions due to the uncertainties in the initial condition of the BK evolution.

\begin{figure}[tb]
    \centering
    \includegraphics[width=1\linewidth]{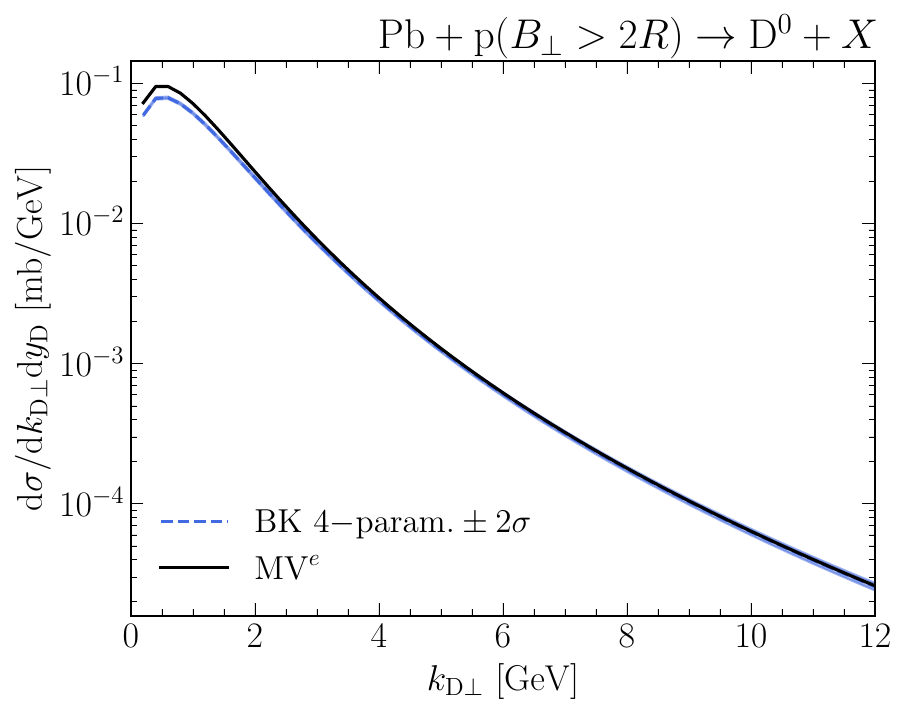}
    \caption{$\dm$ photoproduction cross section at $y_{\mathrm{D}}=1$ and $\sqrt{s}=5.36\ \mathrm{TeV}$ calculated using the BK 4--parametrization of the dipole initial condition in Ref.~\cite{Casuga:2023dcf} with $2\sigma$ uncertainty estimate (blue band). The black line is calculated for proton targets in the MV$^e$ model of Ref.~\cite{Lappi:2013zma}.}
    \label{fig:uncertaintyestimate}
\end{figure}

In Fig.~\ref{fig:uncertaintyestimate} we present the $\dm$ photoproduction cross section with an uncertainty estimate of the BK initial condition in Eq.~\eqref{eq: initial condition dipole} using the BK 4-parameter fit of Ref.~\cite{Casuga:2023dcf}. We only take the dipole-proton scattering amplitude from Ref.~\cite{Casuga:2023dcf}, and as such compute $\dm$ photoproduction in $\gamma+p$ scattering. For the photon flux we use the same as in $\mathrm{Pb}+\mathrm{Pb}$ UPCs, which corresponds to setting $B_\perp>2R$. This is the same as the reference calculated for the nuclear modification factor $R_{pA}$ in Sec.~\ref{sec:results}, and although it does not correspond to any measurable quantity, it provides us a convenient tool to estimate the uncertainty originating from the non-perturbative initial condition of the BK equation. For comparison, in Fig.~\ref{fig:uncertaintyestimate} we also show the $\dm$ photoproduction in the $\gamma+p$ scattering using the MV$^e$ fit \cite{Lappi:2013zma}. We find that the two fits result in similar predictions for $\dm$ photoproduction, with a small deviation ranging from $\sim 10\% $ at low $\kdm$ to $\sim 1\%$ at higher $\kdm$. The $2\sigma$ uncertainty band obtained for our predictions is small, $\sim1\%$. Therefore, we conclude that the dominant source of uncertainty in our calculation is not from the uncertainties in the values obtained for the fit parameters. However, we emphasize that both fits applied here only include light quarks, and including a charm quark in the fit will likely have a significantly larger effect than the uncertainty quantified in this Appendix.

\bibliographystyle{JHEP-2modlong}
\bibliography{refs}
\end{document}